\documentstyle[12pt]{article}
\def\Journal#1#2#3#4{{#1} {\bf #2}, #3 (#4)}

\def\be{\begin{equation}}
\def\ee{\end{equation}}
\def\bea{\begin{eqnarray}}
\def\eea{\end{eqnarray}}

\def\verho{\vec{\rho}} 
\def\barho{\bar{\rho}}   
\def\velambda{\vec{\lambda}}

\def\duasli{'\negthinspace '}
\begin{document}
\title{\bf MASS EFFECTS IN A THREE-BODY SYSTEM BOUND BY HARMONIC OSCILLATORS}
\vspace{2cm}
\author{\it A. C. B. ANTUNES\\
\it Instituto de F\'{\i}sica, Universidade 
 Federal do Rio de Janeiro\\ \it P.O.Box 68528, 
Ilha do Fund\~{a}o, 21945 Rio de Janeiro, RJ, Brazil\\
\it e-mail: Antunes@if.ufrj.br\\
\\
\it L.J. ANTUNES\\
\it Instituto de Engenharia Nuclear - 
CNEN\\
\it P.O.Box 68550, Ilha do Fund\~{a}o, Rio de Janeiro, RJ, Brazil}

\date{}
\maketitle
\vspace{2cm}
\begin{abstract}
The problem of three different masses bound by  harmonic oscillator potentials is solved exactly. It is shown that Jacobi coordinates cannot, in general, decouple this system into two three-dimensional oscillators but this decoupling can always be obtained in terms of the normal coordinates.  The condition for the decoupling in Jacobi coordinates is given.  It is shown that the mean distance between each pair of particles depends upon their masses.
\end{abstract}

\newpage

\setcounter{equation}{0}

\vskip .5 cm 

In this paper we analyze the general problem of three particles with different masses bound by  oscillator potentials with different elastic constants. This problem cannot be decoupled into six independent oscillators in the Jacobi coordinates, as is sometimes assumed. We give the condition that must be fulfilled to decouple that system in Jacobi coordinates. We also give the general solution of the problem in terms of the normal coordinates.

A generic three-particle system has masses $ m_i \enspace (i = 1, 2, 3)$ and positions $\verho_i \enspace(i = 1, 2, 3)$ relative to the overall center of mass. Relative positions are
\begin{equation}
\label{1}
\vec {\rho}_i=\vec{r}_j-\vec{r}_k
\end{equation}
and the position of the particle (i) relative to the center of mass of the pair (jk) is 
\be
\label{2}
\velambda_i=\vec{r}_i-\frac{m_j  \vec{r}_j + m_k  \vec{r}_k}{ m_j + m_k}
\ee
\noindent with (i, j, k) = (1, 2, 3) and cyclic permutations. The kinetic energy can be written in terms of Jacobi coordinates\cite{moshi} as
\be
\label{3}
T=\frac{1}{2}\mu_{jk} (\dot{\verho}_i)^2 + \frac{1}{2}\mu_i (\dot{\velambda}_i)^2
\ee
where
\be
\label{4}
   \mu_{jk}=\frac{m_j m_k}{m_j+m_k} \mbox{\qquad and \qquad} \mu_i=\frac{m_i(m_j+m_k)}{m_1+m_2+m_3}
\ee

The potential energy is
\be
\label{5}
V=\frac{1}{2}{K_i}'{\verho_i}^2+\frac{1}{2}K_{jk}{\velambda_i}^2+{K_i}{'\negthinspace '}\verho_i \cdot \velambda_i 
\ee
\noindent where
\begin{eqnarray}
\label{6}
{K_i}'=K_i+(\mu_{jk}/m_k)^2K_j+(\mu_{jk}/m_j)^2 K_k \nonumber \\
K_{jk}=K_j+K_k \mbox{\enspace, \qquad and} \\
K_i\duasli=\mu_{jk}(K_j/m_k-K_k/m_j) \nonumber
\end{eqnarray}
The third term in the potential energy shows that  the coordinates $\verho_i$ and $\velambda_i$ do not oscillate independently, in general. This only happens if $m_jK_j=m_kK_k$.

We can obtain another set of coordinates in which any three-body system bound by oscillators can be decoupled into two three-dimensional oscillators. For this we use the formalism of small oscillations\cite{gold}, in which the matrices of kinetic and potential energies written in terms of the Jacobi coordinates are simultaneously diagonalized.

The matrices of the kinetic and potential energies are:
\be
\label{7}
T=\left(
\begin{array}{cc}
\mu_{23} & 0 \\
0 & \mu_1 \\
\end{array}
\right)
\quad \mbox{and} \quad
V=\left(
\begin{array}{cc}
K' & K \duasli \\
K \duasli & K_{23}\\
\end{array}
\right) \quad .
\ee
The normal frequencies of the system are given by:
\be
\label{8}
\omega_{1,2}^2=\frac{(\mu_1K'+\mu_{23}K_{23})\pm [(\mu_1K'-\mu_{23}K_{23})^2+4\mu_1\mu_{23}K\duasli ^2]^{1/2}}
{2\mu_1\mu_{23} }
\ee
The corresponding eigenvectors $\vec{a}_l \thinspace \thinspace(l=1,2)$, satisfying the orthonormalization condition are
\bea
\label{9}
\vec{a}_1=[\mu_{23}(K_{23}-\mu_1\omega_1^2)^2+\mu_1K\duasli^2]^{-1/2}
\left(
\begin{array}{cc}
K_{23}-\mu_1\omega_1^2\\
-K\duasli\\
\end{array}
\right) \nonumber\\
\mbox{and} \qquad \qquad \qquad \qquad \qquad \qquad \qquad \\
\vec {a}_2=[\mu_1(K'-\mu_{23}\omega_2^2)^2+\mu_{23}K\duasli^2]^{-1/2}
\left(
\begin{array}{cc}
-K\duasli \\
K'-\mu_{23}\omega_2^2\\
\end{array}
\right) \nonumber 
\eea

    The matrix that diagonalizes simultaneously the kinetic and potential energy matrices T and V is  
\be
\label{10}
A=\left(
\begin{array}{cc}
\frac{K_{23}-\mu_1\omega_1^2}{[\mu_{23}(K_{23}-\mu_1\omega_1^2)^2+\mu_1K\duasli^2]^{1/2}} & \frac{-K\duasli} {[\mu_1(K'-\mu_{23}\omega_2^2)^2+\mu_{23}K\duasli^2]^{1/2}}\\
\frac{-K\duasli}{[\mu_{23}(K_{23}-\mu_1\omega_1^2)^2+\mu_1K\duasli^2]^{1/2}} &
\frac{K'-\mu_{23}\omega_2^2}{[\mu_1(K'-\mu_{23}\omega_2^2)^2+\mu_{23}K\duasli^2]^{1/2}}\\
\end{array}
\right)
\ee                            
The normal coordinates of the three-particle system are defined by
\be
\label{11}
\left(
\begin{array}{cc}
\rho_{1\alpha}\\
\lambda_{1\alpha}\\
\end{array}
\right)
=A\left(
\begin{array}{cc}
\xi_{1\alpha}\\
\xi_{1\alpha}\\
\end{array}
\right) \qquad \qquad (\alpha=x,y,z)
\ee
This relation may be inverted using the orthonormality condition \~{A}TA=I and the normal coordinates are given by
\bea
\label{12}
\vec{\xi_1}=\frac
{\mu_{23}(K_{23}-\mu_1\omega_1^2)\verho_1-\mu_1K\duasli\velambda_1}
{[\mu_{23}(K_{23}-\mu_1\omega_1^2)^2+\mu_1K\duasli^2]^{1/2}} \nonumber \\
\quad \\
\vec {\xi_2}=\frac
{-\mu_{23}K\duasli\verho_1+\mu_1(K'-\mu_{23}\omega_2^2)\velambda_1}
{[\mu_1(K'-\mu_{23}\omega_2^2)^2+\mu_{23}K\duasli^2]^{1/2} } \nonumber
\eea
   The conjugate momentum of the coordinate $\xi_{\alpha j}$ is $\pi_{\alpha j}=\partial L/\partial \dot{\xi}_{\alpha j}= \dot{\xi}_{\alpha j}$, and the Hamiltonian of the system, corresponding to six independent oscillators, is
\be
\label{13}
H=\sum_{\alpha=x,y,z}\thinspace \sum_{j=1,2}\frac{1}{2}(\pi_{\alpha j}^2+\omega_j^2 \xi_{\alpha j}^2)
\ee       
    The Schroedinger equation $H_j\psi(\vec {\xi}_j)=E_j\psi(\vec{\xi}_j)$ can be solved in the same way as the three-dimensional harmonic oscillator \cite{powe,messiah}.\\
  
MEAN DISTANCES IN THE GROUND-STATE\\

    For simplicity we choose a particular case in which $K_i = K \enspace (i=1,2,3)$  and $m_3=m_2$. The unnormalized wave function of the ground-state can be written as
\[
\psi_0=e^{-[(\verho_1/\barho_1)^2+(\verho_2/\barho_2)^2+(\verho_2/\barho_3)^2]}\mbox{\qquad \qquad where}  
\]
\be
\frac{1}{\barho_1^2}=\frac{\sqrt{m_2K}}{2\hbar(m_1+2m_2)}\left[m_2\sqrt{3}+\frac{1}{2}\sqrt{m_1}\left(\sqrt{3m_1}-\sqrt{m_1+2m_2}\right)\right]
\ee
\be
\mbox{and \qquad \qquad}
\frac{1}{\barho_2^2}=\frac{1}{\barho_3^2}=\frac{\sqrt{Km_1m_2(m_1+m_2)}}{2\hbar(m_1+2m_2)}
\ee
The mean distances between the three particles form a triangle whose proportions are given by $\barho_2=\barho_3$ and
\be
\frac{\barho_2}{\barho_1}=\left[  \frac{\sqrt{3}\frac{m_2}{m_1}+\frac{1}{2}\left(\sqrt{3}-\sqrt{1+2\frac{m_2}{m_1}}\right)} {\sqrt{1+2\frac{m_2}{m_1}}}\right]^{1/2}
\ee

The three possibilities are :  $m_1 \le m_2 = m_3$ then $\barho_1 \le \barho_2 = \barho_3$, and $m_1>m_2=m_3$, that gives $0.605 < (\barho_2 / \barho_1)<1$.

In the limit $\frac{m_2}{m_1}\longrightarrow0$ we have $\barho_2=\barho_3=0.605\barho_1$, and the angle between $\barho_2$ and  $\barho_3$ is  $\alpha=2sen^{-1}(\barho_1 / 2\barho_2)=111.47^{\circ}$.

     These results show that a three-particle system bound by equal oscillators has a spatial distribution that depends on the masses of the particles.  The mean distance between a pair of particles decreases with increasing masses of the pair.

This phenomenon must be taken into account in the construction of certain approximate models for three-body problems, as the diquark model for baryons\cite{skytt}.

In the potential model for baryons, the three-valence quarks interactions are described by the same potential.  This potential is somewhat more complicated than the harmonic oscillator one, and the three-body problem cannot be solved exactly.

A phenomenological approach is the so called diquark model, which is based on evidences that each pair of quarks in the baryon tend to form a cluster that interacts as a whole with the third quark\cite{anselm}.  Our results suggest that in applying the diquark model for baryons with quarks of different masses, the three possible diquarks are not equally probable.  Similarly to the oscillator model, in the realistic potential model the dimension of the diquark cluster depend on the masses of the constituent quarks.  The dimension decrease with increasing masses , then the diquark becomes best defined inside the baryon.

\end{document}